\documentclass[12pt,preprint]{emulateapj}

\usepackage{graphicx}
\usepackage{amsmath}

\begin{document}

\title{Anomalies in Time Delays of Lensed Gravitational Waves and Dark Matter Substructures}
\author{Kai Liao$^{1}$, Xuheng Ding$^{2}$, Marek Biesiada$^{3,4}$, Xi-Long Fan$^{5}$ Zong-Hong Zhu$^{2,3}$}
\affil{
$^1$ {School of Science, Wuhan University of Technology, Wuhan 430070, China.}\\
$^2$ {School of Physics and Technology, Wuhan University, Wuhan 430072, China.}\\
$^3$ {Department of Astronomy, Beijing Normal University, Beijing 100875, China.}\\
$^4$ {Department of Astrophysics and Cosmology, Institute of Physics, University of Silesia, 75 Pu{\l}ku Piechoty 1, 41-500 Chorz{\'o}w, Poland.}\\
$^5$ {Department of Physics and Mechanical and Electrical Engineering, Hubei University of Education, Wuhan 430205, China.}
}
\email{liaokai@whut.edu.cn}

\begin{abstract}
The cold dark matter scenario of hierarchical large-scale structure formation predicts the existence of abundant subhalos around large galaxies.
However, the number of observed dwarf galaxies is far from this theoretical prediction, suggesting that most
of the subhalos could be dark or quite faint. Gravitational lensing is a powerful tool to probe the mass distribution directly irrespective of whether it is visible or dark.
Time delay anomalies in strongly lensed quasar systems
are complementary to flux ratio anomalies in probing dark matter substructure in galaxies.
Here we propose that lensed gravitational waves detected by the third-generation ground detectors
with quite accurate time delay measurements could be a much better tool for this study than conventional techniques.
Combined with good quality images of lensed host galaxies identified by the electromagnetic counterpart measurements, lensed GW signals could make the systematic errors caused by dark matter substructures detectable at several percent levels,
depending on their mass functions, internal distribution of subhalos and  lensing system configuration.

\end{abstract}

\section{Introduction}
The cold dark matter (CDM) model predicts that about $25\%$ of the matter content in the Universe is of non-baryonic origin and
large dark matter halos have been assembled hierarchically from smaller ones. While this model has successfully
explained the large-scale structure of the Universe at the level of galaxies and galaxy clusters, its
sub-galactic scale predictions have not yet been well tested. According to the simulations, a small part of
galactic dark matter halos should be in the form of subhalos that temporarily
survived from tidal stripping process. The mass function of these clumps approximately follows a power law function $dN/dm\propto m^{-1.8}$~\citep{Diemand2008}. It is believed that subhalos anchoring gas allowing for star formation would
appear as satellite dwarf galaxies. Therefore, we are supposed to observe a lot of such satellites.
However, the long-standing ``Missing Satellite Problem" makes the picture blurred. The CDM simulations predict that thousands of subhalos should be bound to the Milky Way~\citep{Klypin1999}, while only $\sim 10$ luminous satellites have been observed~\citep{Drlica2015}. The same is true in the Andromeda M31 galaxy~\citep{Moore1999}.

To solve the mismatch between the low-mass end of the subhalo mass function and the luminosity
function of dwarf galaxies, various mechanisms were proposed. For example, processes inhibiting star formation in low-mass subhalos, or observational biases that rule faint satellites out of the surveys.
Another possibility is that substantial scatter exists among galaxies, i.e., the Milky Way and Andromeda are quite special
ones. Another approach was taken by theorists, who tried to modify dark matter properties to decrease the formation of low-mass subhalos. These ideas include: warm dark matter~\citep{Lovell2014}, self-interacting dark matter~\citep{Spergel2000}, fuzzy dark matter~\citep{Wayne2000} and superWIMPs~\citep{Land2005}. Inflation was also adjusted to reduce the low-mass end of sub-halos~\citep{Kamionkowski2000}. Therefore, measuring the subhalo mass function and how does it vary with the environment are quite important questions, essential for both basic physics and astrophysics.

Strong gravitational lensing is an excellent tool to directly detect substructure in galaxies outside the local group~\citep{Mao1998,Zackrisson2010}, since
it does not distinguish between luminous and dark matter. Lensed quasar systems have been used
to detect dark matter substructure using the observed flux ratio anomalies.
In many cases, while the smooth lens model can fit the image positions well, yet flux ratios among images
become anomalous most probably due to the substructure in the lensing galaxy dark matter halos. The smooth model here, can not be
formulated non-parametrically in terms of the multipole expansion, since in such case it would lead to unrealistic galaxy shapes~\citep{Keeton2009}.
On the other hand, with a parametric smooth model, one can infer the properties of subhalos. However, this approach may suffer from propagation effects in the interstellar medium~\citep{Mittal2007} and microlensing effects by the motion of stars in the lensing galaxy~\citep{Schechter2002}. Selecting different wavelengths can effectively mitigate these biases~\citep{Jackson2015,Nierenberg2017}. Besides, the astrometric effects~\citep{Koopmans2002,Vegetti2009} or the
small-scale structure in macro-images~\citep{Inoue2005} could also be utilized
to study dark matter substructure.

Keeton and Moustakas~\citep{Keeton2009} proposed that time delay perturbations between macro-images could complement the methods mentioned above by measuring different moments of the substructure mass function and applied it to well studied lensed quasar systems RX J1131-1231 and B1422+231~\citep{Congdon2010}. This approach is immune to dust extinction or stellar microlensing.  Note that image positions and flux ratios depend on the first and the
second derivatives of the lens potential, respectively, while time delay depends directly on the lens potential.

To achieve a robust identification of time delay anomalies, one needs to simultaneously improve accuracies of both the
smooth model and the measurements of time delays.
However, the light curves of lensed quasars allow for at most
$\sim 3\%$ accuracy of time delays~\citep{Liao2015}, while the typical perturbation is only a fraction of a day~\citep{Keeton2009}.
Recently, \citep{Tie2018}
suggested that a new microlensing effect on time delays based on differential magnification of the
accretion disc of the lensed quasar, may further increase the uncertainties up to $30\%$.
Besides, smooth model uncertainties in lensed quasars were based on the assumption of point sources and Monte Carlo simulation based on specific galaxy catalogs~\citep{Congdon2010}. Therefore, they   could mask the time delay perturbations.
Consequently, one might doubt, whether the anomalies found in RX J1131-1231 and B1422+231 were directly related with dark matter subhalos.

Recent detections, by the Advanced Laser Interferometer Gravitational Wave Observatory (adLIGO), of gravitational wave (GW) signals generated in mergers of binary black holes (BHs) opened a new window on the Universe ~\citep{GW150914,GW151226,GW170104}. Lensing of GW by intervening masses (galaxies) have been discussed by ~\citep{Wang1996,Nakamura1998,Takahashi2003,Cao2014,Sereno2010}.
Furthermore, the observed electromagnetic (EM) counterpart of the binary neutron stars merger opened a new chapter in the multi-messenger astronomy~\citep{GW170817}.
The next generation of GW interferometric detectors, like the Einstein Telescope (ET) will broaden the accessible volume of the Universe by three orders of magnitude with forecasted tens to hundreds of thousands of detections per year~\citep{Abernathy2011} leading to expectation that many of the sources could be gravitationally lensed. This was discussed by~\citep{JCAP_ET1,JCAP_ET2,JCAP_ET3} with a conclusion that ET should register about 50 -- 100 strongly lensed inspiral events per year, thus providing a considerable catalog of such events during a few years of its successful operation.
Lensed GW signals accompanied by EM counterparts are supposed to be valuable in the context of both the fundamental physics~\citep{Fan2017} and cosmology~\citep{Liao2017}.

In this paper, we propose to use lensed gravitational wave signals together with their electromagnetic counterparts to identify time delay anomalies and study dark matter substructure. \textbf{For simplicity, we attribute all anomalies to the dark matter substructure. However, we emphasize that for some realistic systems,
complex baryonic structure can also contribute to the observed anomalies~\citep{Hsueh2016,Hsueh2017}. Ignoring the full complexity of the lens macro-model
would overestimate the dark matter substructure component~\citep{Xu2015,Gilman2017,Evans2003}.}
Despite of this, our analysis shows that
systematic uncertainties caused by dark matter subhalo perturbations can be detected to some percent levels,
depending on the subhalo mass function, internal structure of subhalos, and lensing system configuration.


\section{Systematics by dark matter substructure}

According to gravitational lensing theory~\citep{Treu2010}, time delay between multiple images $i, j$ is given by:
\begin{equation}
\Delta t_{i,j} = \frac{D_{\mathrm{\Delta t}}(1+z_{\mathrm{d}})}{c}\Delta \phi_{i,j}, \label{relation}
\end{equation}
where $c$ is the speed of light, $\Delta\phi_{i,j}=[(\boldsymbol{\theta}_i-\boldsymbol{\beta})^2/2-\psi(\boldsymbol{\theta}_i)-(\boldsymbol{\theta}_j-\boldsymbol{\beta})^2/2+\psi(\boldsymbol{\theta}_j)]$
is the Fermat potential difference for image angular positions $\boldsymbol{\theta}_i$ and $\boldsymbol{\theta}_j$; $\boldsymbol{\beta}$ denotes the source position, and $\psi$ is the two-dimensional lensing potential determined by the surface mass density of the lens
$\kappa$ in units of critical density $\Sigma_{\mathrm{crit}}=c^2D_{\mathrm{s}}/(4\pi GD_{\mathrm{d}}D_{\mathrm{ds}})$ through the Poisson equation $\nabla^2\psi=2\kappa$,
 $D_{\mathrm{d}}$, $D_{\mathrm{s}}$ and $D_{\mathrm{ds}}$ are angular diameter distances to the lens (deflector) located at redshift $z_{\mathrm{d}}$, to the source located at redshift $z_{\mathrm{s}}$ and between them, respectively. Dark matter substructure could perturb Fermat potentials including gravitational potential and image positions
 and therefore could perturb time delays.

Lensed GW signals accompanied by electromagnetic counterparts, in particular kilonovae that are relatively stable and easy to observe~\citep{counterparts}, are especially advantageous for
studying the dark matter substructure. Firstly, time delay measured by GWs could be quite
accurate due to the transient nature of the event. They can be determined with accuracy $\sim 0.1s$~\citep{Fan2017}, and such measurement is essentially waveform independent.
Secondly, the kilonovae last only for months, so
one could measure the entire host galaxy arcs before or after the electromagnetic counterpart, which strongly facilitates
lens modelling~\citep{Liao2017}. In addition to the analysis of ~\citep{Keeton2009}, 
time delay measurements should be not affected by microlensing due to the long wavelengths of GWs in the diffraction limit~\citep{Takahashi2003}.

Time delays of lensed GW signals observed together with their electromagnetic counterparts are affected by
at least five types of uncertainties. First, is the combined observational uncertainty $\sigma_{obs}$ comprising pixel intensities, central velocity dispersion and point image positions. Time delay measurements are not included since we assume they would be determined accurately with the GW signals. This component can be thought of as the smooth lens model uncertainty. Next is  $\sigma_{LOS}$ arising from
the mass density fluctuation along the line of sight. Then, the uncertainty $\sigma_{cosm}$ stemming form the cosmological model adopted in calculations of distances should be taken into account. It captures a  possible mismatch between the true and fiducial background cosmological model. The fourth component $\sigma_{dm}$ is directly caused by dark matter subhalo perturbations. Note that astrometric effects are included here because we measure time delays of the perturbed images.
The last component is $\sigma_{arc}$ due to the perturbation of images (arcs) by dark matter halos. This systematic component would in turn affect the accuracy of the smooth lens model.
The total uncertainty is given by:
\begin{equation}
\sigma_{tot}^2=\underbrace{\sigma_{obs}^2+\sigma_{LOS}^2+\sigma_{cosm}^2}_
 {\sigma_{stat}^2}+\underbrace{\sigma_{arc}^2+\sigma_{dm}^2}_{\sigma_{sys}^2},\label{uncertainties}
\end{equation}
$\sigma_{tot}$ can been seen as the difference between the measured time delay and the one inferred from the smooth lens model best fitted to images. From the perspective of dark matter substructure we propose to treat the first three terms collectively as statistical uncertainties, even though some of them are in fact systematics (but due to effects different from the one we are focused on).
In this convention, the last two components are systematical ones --- caused by dark matter halos resulting in apparent anomalies. Note that $\sigma_{dm}$ is correlated with $\sigma_{arc}$. However, as
we will discuss later, this correlation could be neglected.

Since we assume that measurements of time delays by GW signals are accurate, the corresponding extra uncertainty is $\sigma_{\Delta t}=0$.  If applied to lensed quasars, one would need to consider additional $3\%$ uncertainty of $\Delta t$ from light curves and moreover $\sigma_{obs}$ would be larger due to the bright AGNs contaminating the arcs.

\textbf{One should also include another systematical uncertainty from the selection of smooth macro model. For example, a power-law model and a composite model give different results~\citep{Wong2017}.
However, this systematics should not exceed the scale of $\sigma_{obs}$ and can be well controlled according to current techniques~\citep{Wong2017, Suyu2013, Suyu2017}. Therefore, we do not include it explicitly in this work assuming that $\sigma_{obs}$ is sufficient. }

\section{Simulations and results}
In order to illustrate our idea, we investigate $\sigma_{dm}, \sigma_{obs}, \sigma_{arc}$ for the fiducial double and quad systems, respectively. The other two uncertainties $\sigma_{LOS}$ and $\sigma_{cosm}$
are estimated from different inputs, so in both cases we take them as
$2\%$ and $1\%$, respectively~\citep{Rusu2017}.

\subsection{Reference lensing system}
\textbf{In principle, a realistic approach capable of revealing more potential systematics should be base on numerical simulations of the lens. However, in order to illustrate our idea we only use a simple macro model.
We refer to the ongoing work~\citep{Ding2018}, where the Time Delay Lens Modelling Challenge (TDLMC) program will thoroughly investigate  systematic errors and biases of the lens model, with the purpose to check whether precision could dominate systematics in current lensing techniques.}

To show how time delay anomalies are related to dark matter substructure, we focus on a specific lensing system with the source and the lens redshifts 1.5 and 0.5, respectively.
The reference smooth lens galaxy is modeled by a singular isothermal ellipsoid (SIE), with three-dimensional radial profile $\rho(r)\propto r^{-2}$,
central velocity dispersion $\sigma_v=300km/s$, ellipticity $e=0.2$ and orientation $\theta_e=45^\circ$.
Besides, we add a constant shear modelling the impact of the lens environment: the amplitude is $0.003$ with orientation $\theta_\gamma=120^\circ$.

\begin{figure*}
\centering
 \includegraphics[width=5cm,angle=0]{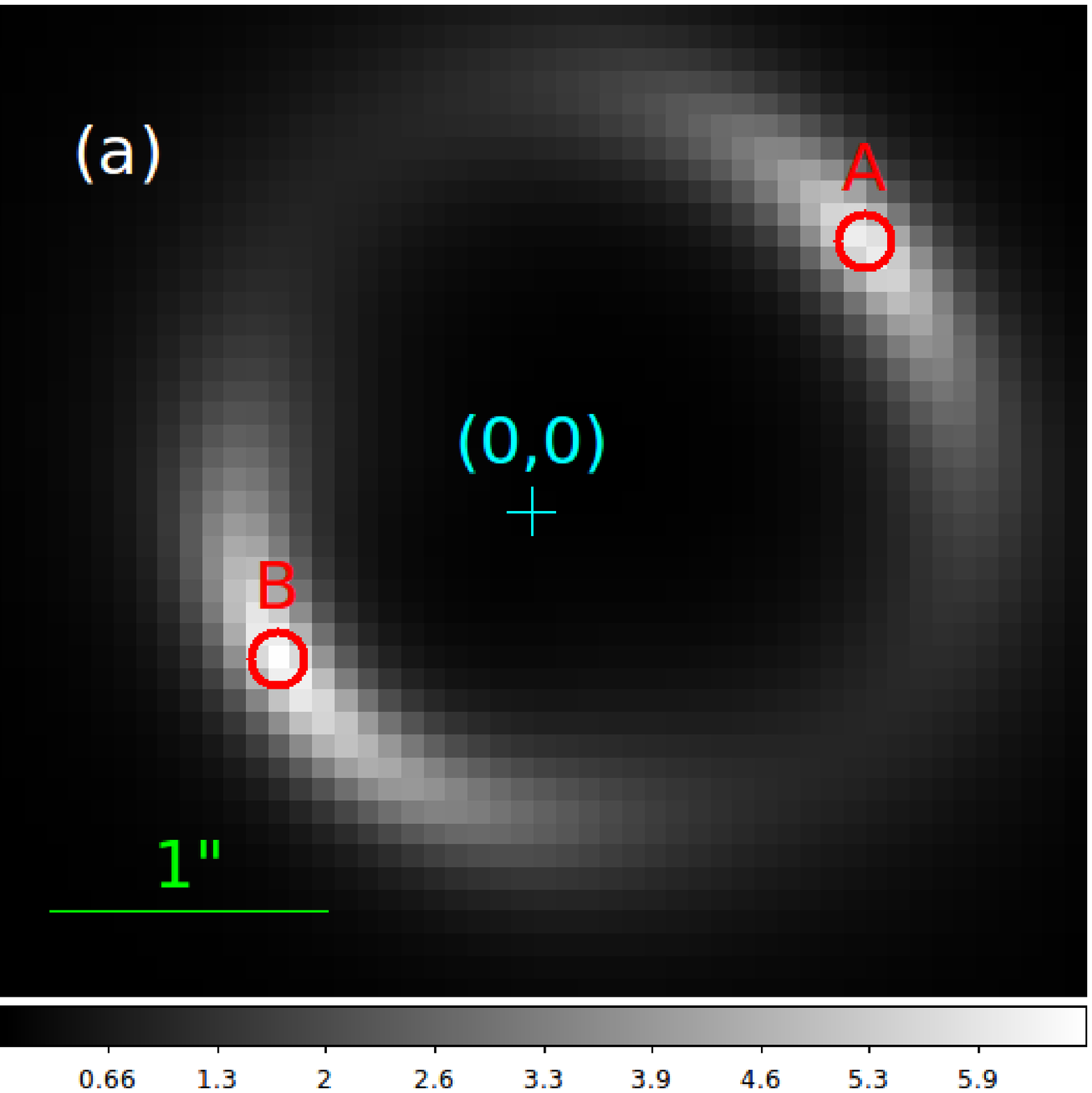}
 \includegraphics[width=5cm,angle=0]{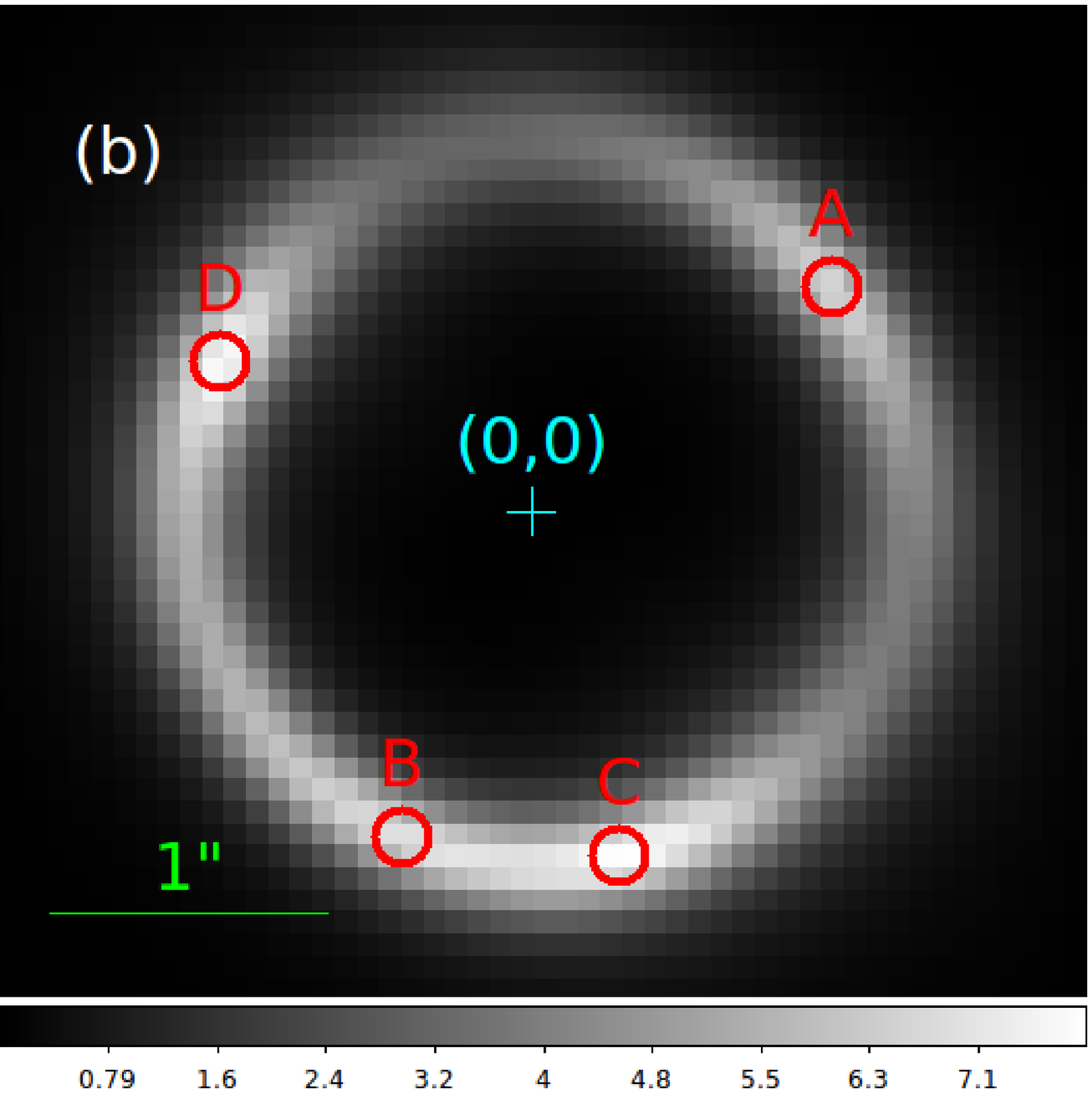}
 \includegraphics[width=5cm,angle=0]{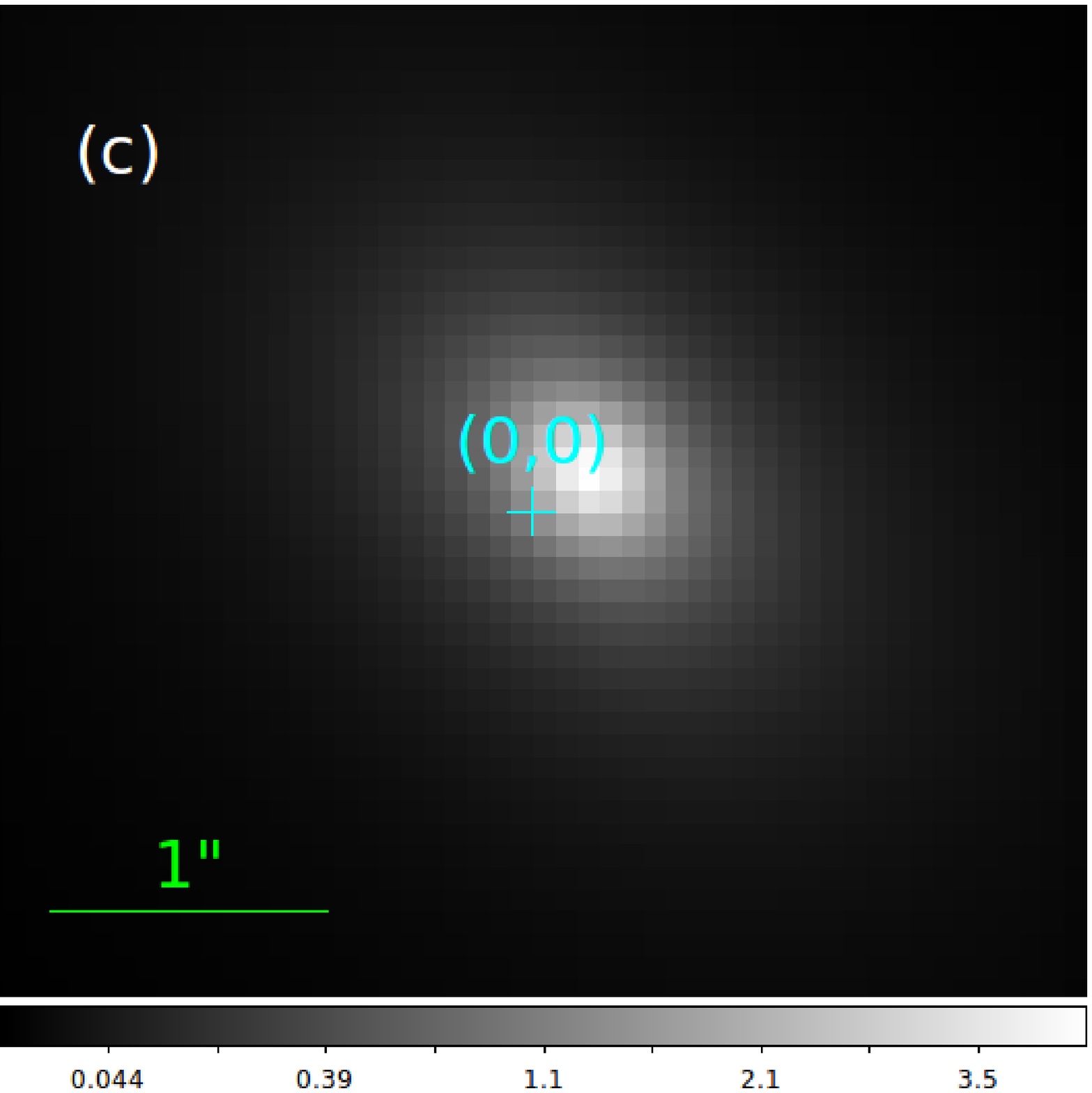}
 \includegraphics[width=5cm,angle=0]{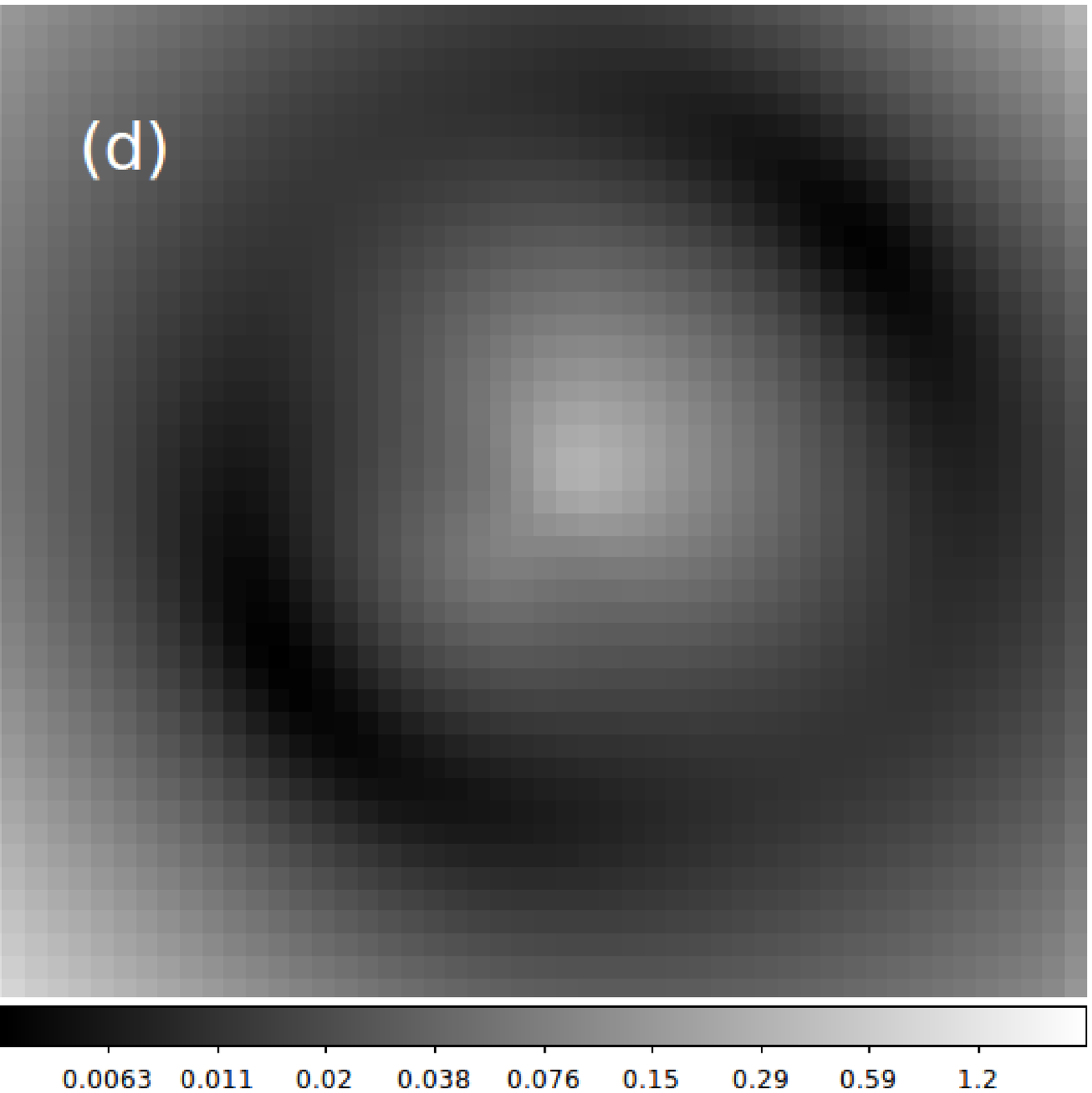}
 \includegraphics[width=5cm,angle=0]{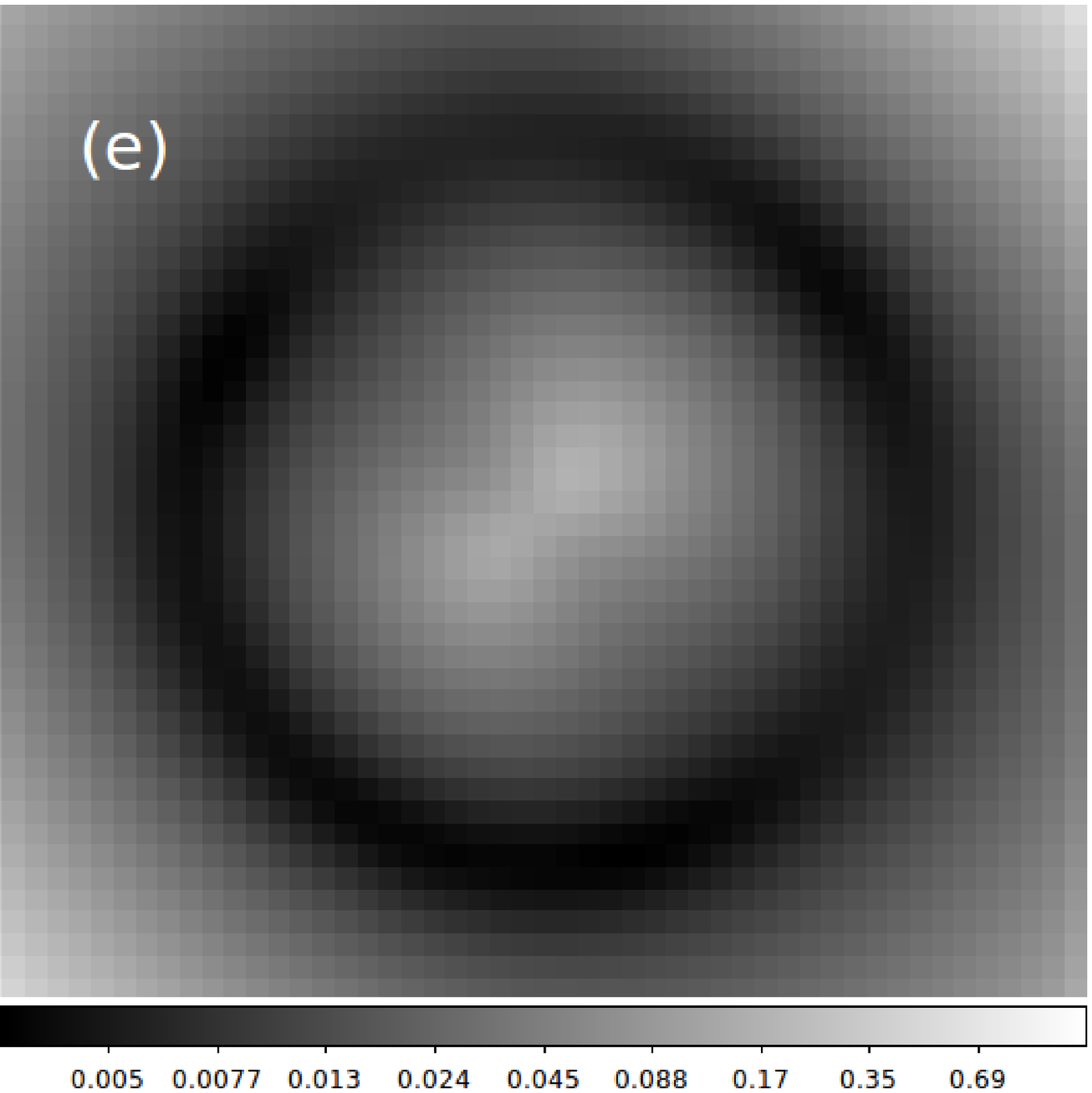}
 \includegraphics[width=5cm,angle=0]{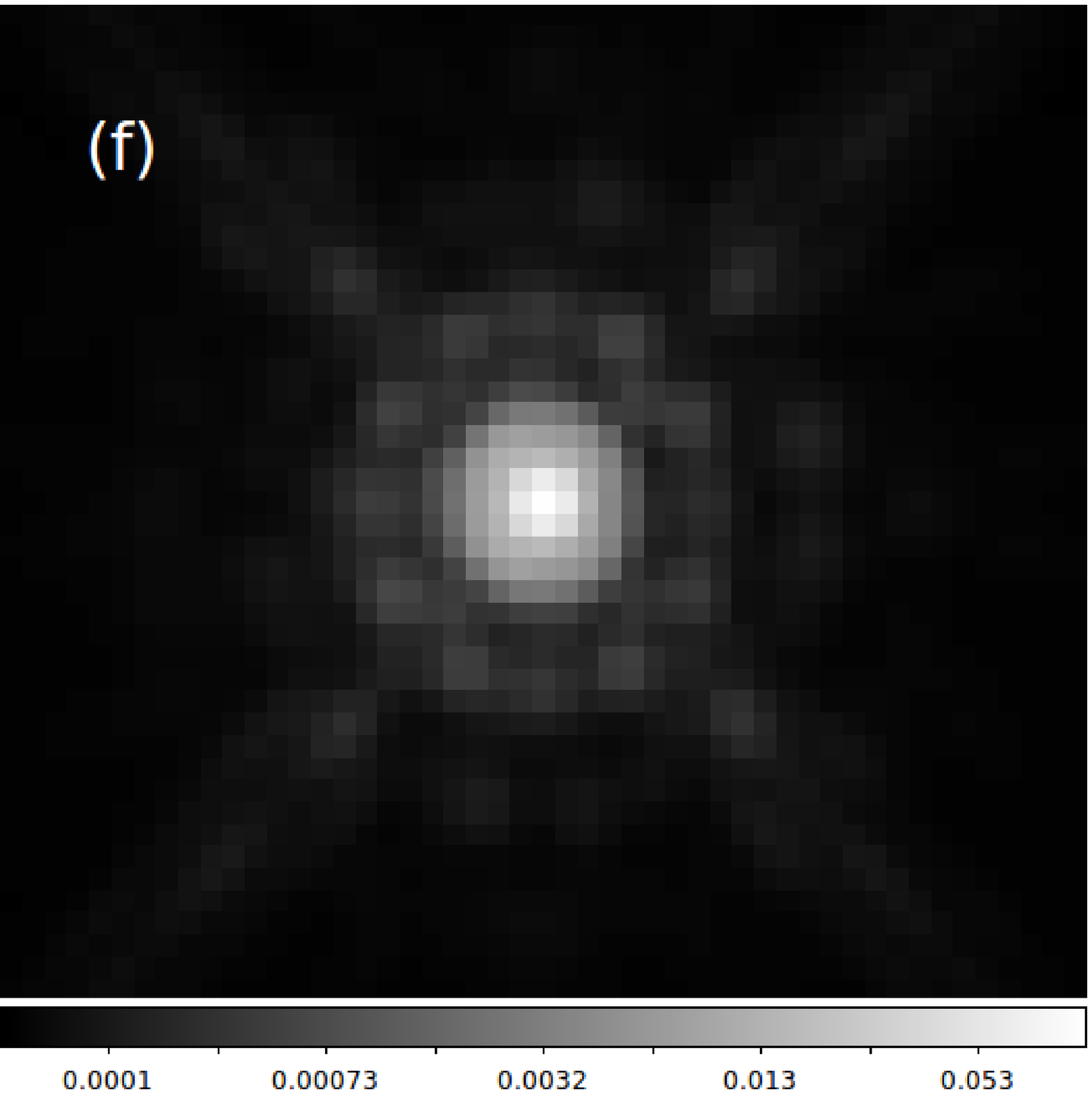}

  \caption{(a) simulated host arc image of double-image system without dark matter subhalo perturbation and noise; (b)
   simulated host arc image of quad-image system without dark matter subhalo perturbation and noise; (c) original source image without lensing;
   (d) relative noise map for double-image system; (e) relative noise map for quad-image system; (f) PSF image based on F160W from TINY TIM.
  }\label{images}
\end{figure*}

The reference host galaxy of the source is modeled by the S$\acute{e}$rsic model, where the projected mass profile is:
\begin{equation}
\Sigma(r)=\frac{M_{tot}}{\pi r_s^2\Gamma(2n+1)}e^{-(\frac{r}{r_s})^{1/n}},
\end{equation}
where $M_{tot}$ is the total mass and is proportional to the total brightness $L_{tot}$, $r_s$ is the scale radius related to the effective radius $r_e$ and the S$\acute{e}$rsic index, $\Gamma(z)$ is the Euler's Gamma function, $n$ is the S$\acute{e}$rsic index controlling the concentration: $n=1$ corresponds to exponential disk, $n=4$ corresponds to de Vaucouleurs profile. We assume $L_{tot}=250, e=0.3, \theta_e=120^\circ, r_e=0.35'', n=2$. The source was put at two positions: $\boldsymbol{\beta}=(0.15'', 0.05'')$ and $\boldsymbol{\beta}=(0.05'',0.05'')$ so that it can form double image and quadruple image systems.
Fig.~\ref{images} (a) (b) (c) shows the lensed host galaxy images and the unlensed one, all are without noise, in one second exposure time.

\subsection{Subhalo population and $\sigma_{dm}$}
According to the CDM simulations, dark matter subhalo mass function approximately follows
a power law, $dN/dM\propto m^\beta$ with $\beta\approx-1.8$. For simplicity and the purpose of illustration, we assume that dark matter subhalos trace the total mass, $\kappa_s(r)=f_s\kappa_{tot}(r)$.
We also consider finite range of subhalo mass. For the case 1, $f_s=0.01$, we choose $10^7M_{sun}<m<10^9M_{sun}$ and assume
the subhalos are modelled by point mass.  For the case 2, $f_s=0.01$,   we assume that the subhalos have internal structure
modelled by pseudo-Jaffe model, the respective parameters of which are shown in Tab~\ref{jaffe}. For case 3, $f_s=0.001$, we choose $10^6M_{sun}<m<10^8M_{sun}$ modelled by point mass. More examples of  realistic mass functions, can be found in~\citep{Keeton2009}.
To illustrate the impact of subhalo mass and dark matter fraction, we considered cases 1 and 2 for double-image system,
and cases 2 and 3 for quad system.

\begin{figure*}
 \includegraphics[width=6cm,angle=0]{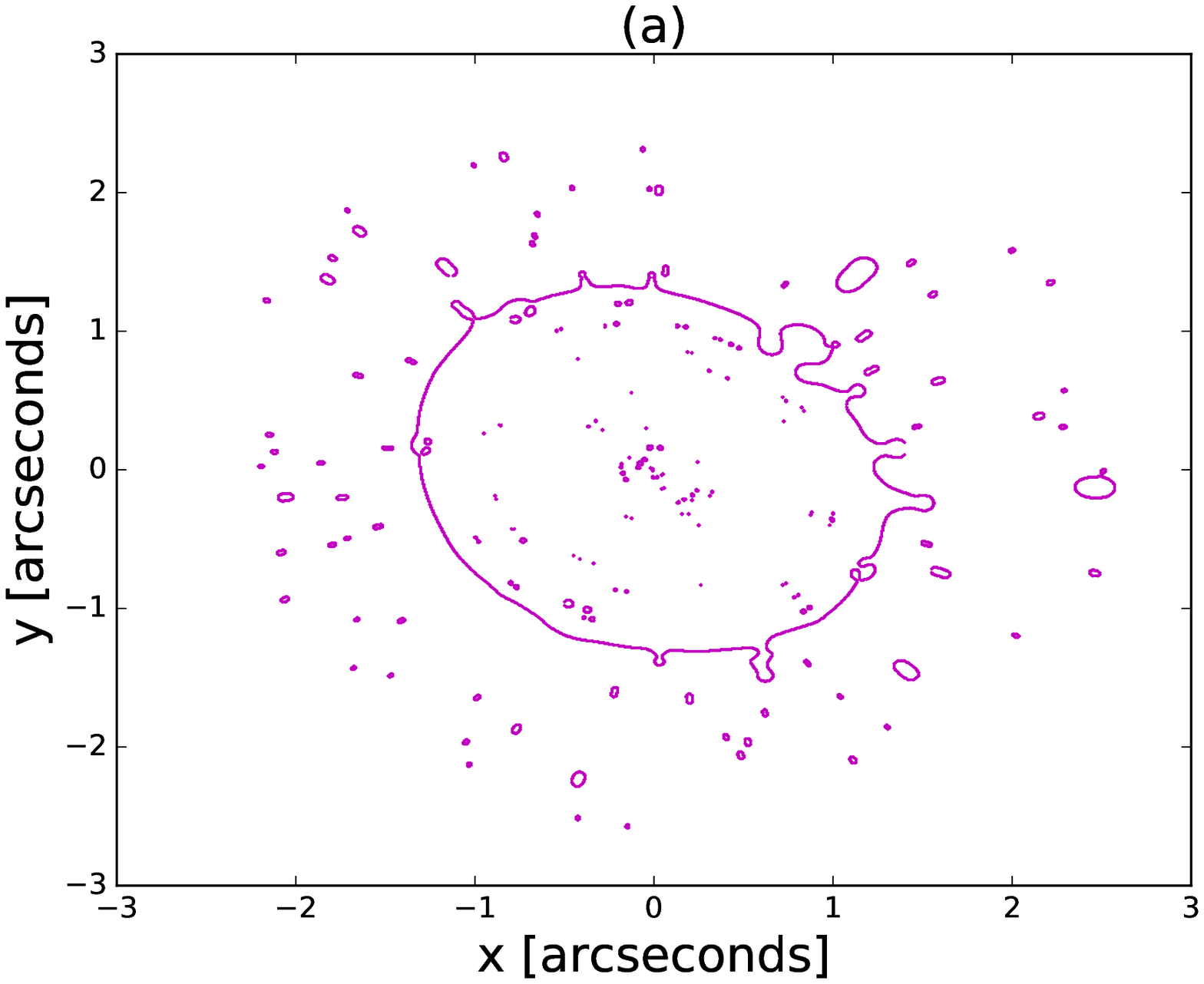}
 \includegraphics[width=6cm,angle=0]{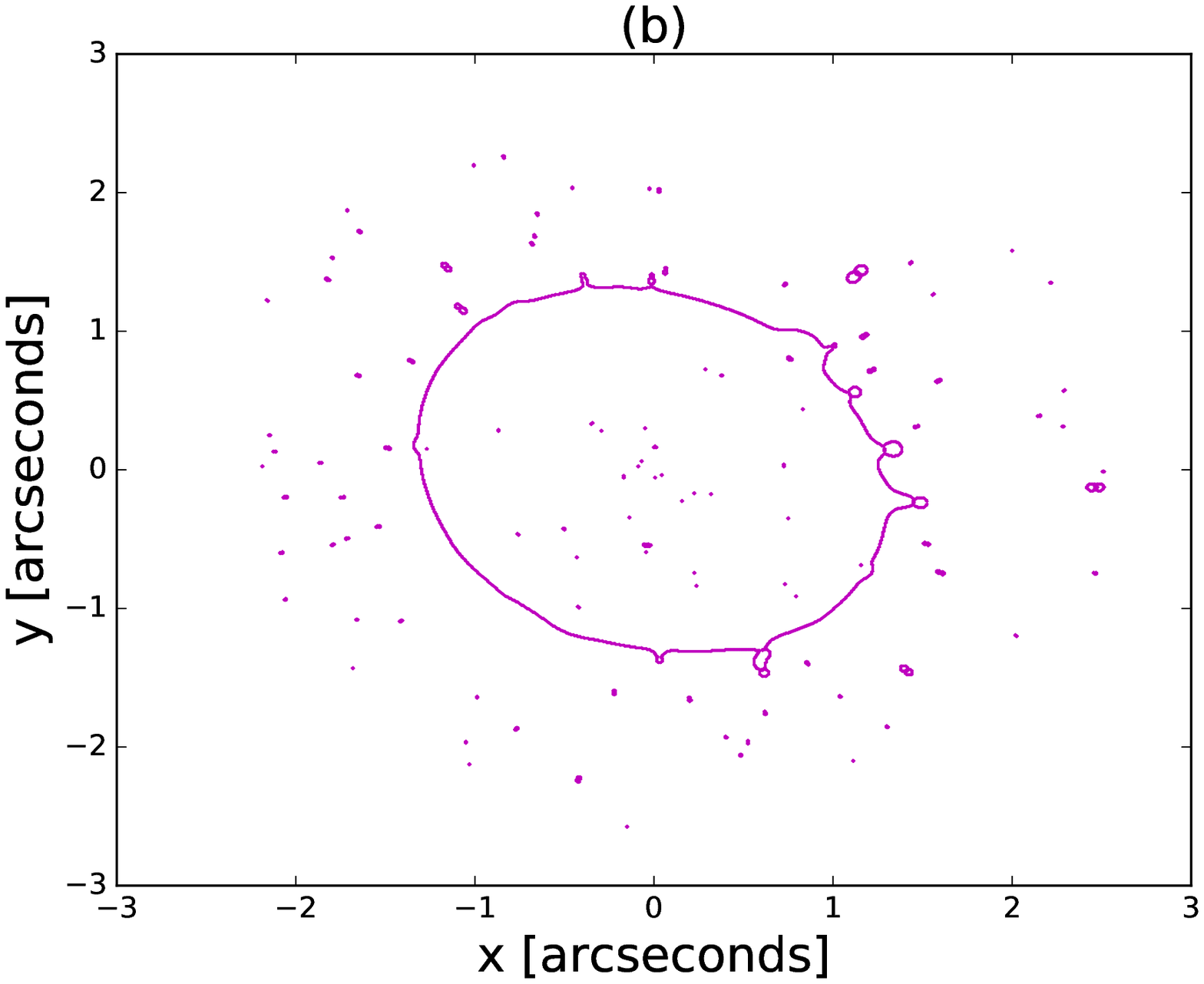}
 \includegraphics[width=6cm,angle=0]{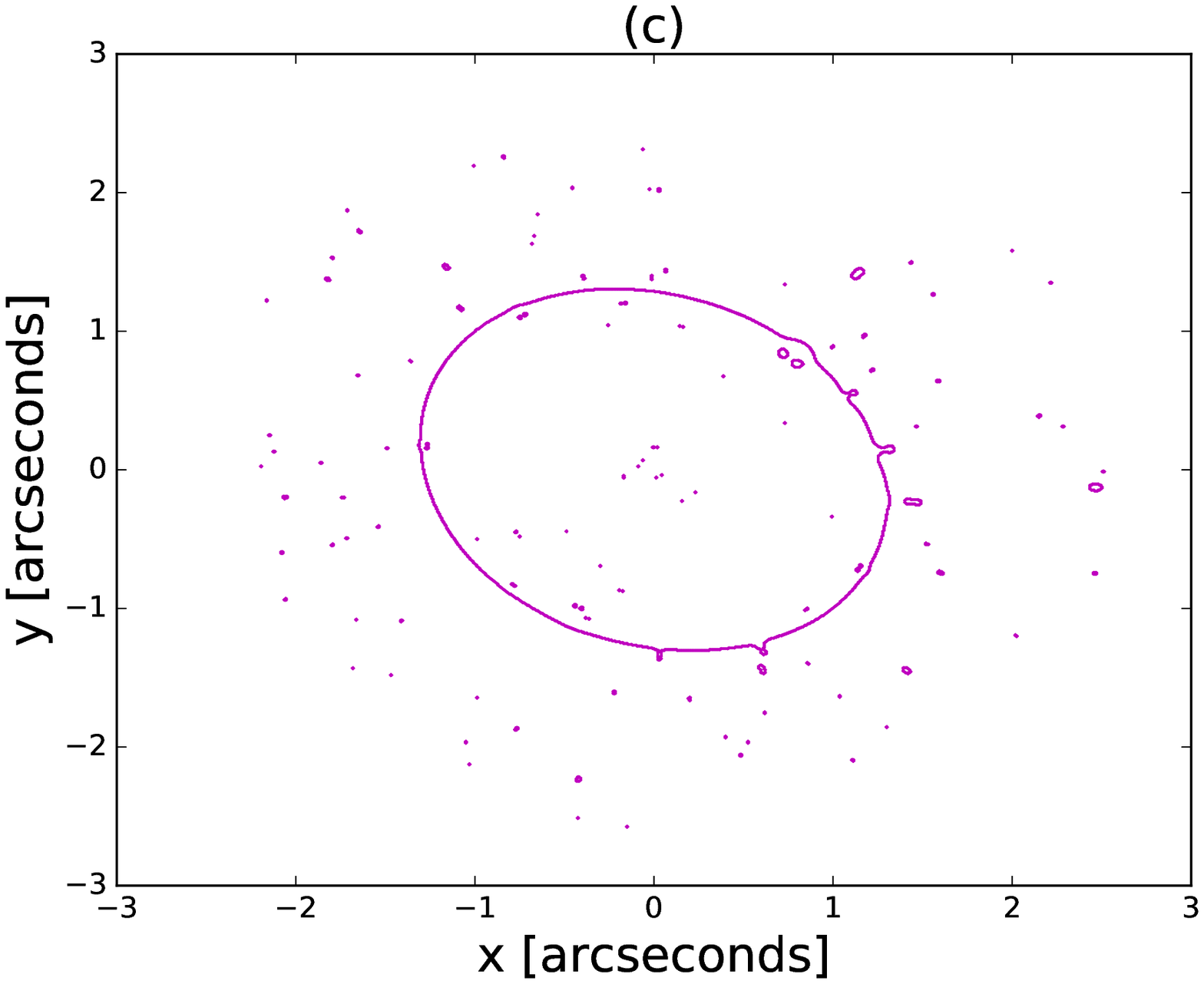}
  \caption{Perturbed critical lines for different mass functions and internal structures. (a) case 1; (b) case 2; (c) case 3.
  } \label{crit}
\end{figure*}

Using publicly available lensing software ``glafic"~\citep{Oguri2010},
we first calculated the distribution of convergence $\kappa(r)$ and then converted it to surface mass density.
We assumed that $f_s$ of the total mass within two Einstein radii is in dark matter. We randomly put these subhalos in the lens plane. Fig. \ref{crit} shows the critical lines for three cases mentioned above, in order to illustrate the impact of dark matter perturbations.

\begin{table*}\centering
 \begin{tabular}{lccccc}
  \hline\hline
  $m/M_{sun}$ &$\sigma (km/s)$& $r_{trun}('')$ & $r_{core}('')$ & $r_{ein}^{jaffe}('')$ & $r_{ein}^{point}('')$\\
  \hline
  7.76e+07  & 19.5 & 0.05 & 0.002 & 0.0034  & 0.0202\\
  4.57e+07  & 19.7 & 0.03 & 0.002 & 0.0032  & 0.0155\\
  2.69e+07  & 17   & 0.024 &0.002 & 0.0012  & 0.0119\\
  1.58e+07  & 15.3 & 0.017 &0.001 & 0.0023  & 0.0091\\
  1.31e+08  & 20   & 0.08 & 0.002 & 0.0039  & 0.0263\\
  2.23e+08  & 26   & 0.08 & 0.002 & 0.0082  & 0.0342\\
  3.80e+08  & 35.1 & 0.08 & 0.007 & 0.0086  & 0.0446\\
  6.45e+08  & 36.4 & 0.123 &0.008 & 0.0092  & 0.0581\\
  1.09e+09  & 46.2 & 0.13  &0.01  & 0.0186  & 0.0757\\
  \hline\hline
 \end{tabular}
 \caption{ Typical parameter values used to model subhalos with pseudo-Jaffe elliptical model $\rho\propto(r^2+r_{core}^2)^{-1}(r^2+r_{trun}^2)^{-1}$, where $m$ is the subhalo mass, $\sigma$ is the central velocity dispersion. We calculated Einstein radii for both the extended  structure and the point mass.
}\label{jaffe}
\end{table*}

\begin{figure}
 \includegraphics[width=8cm,angle=0]{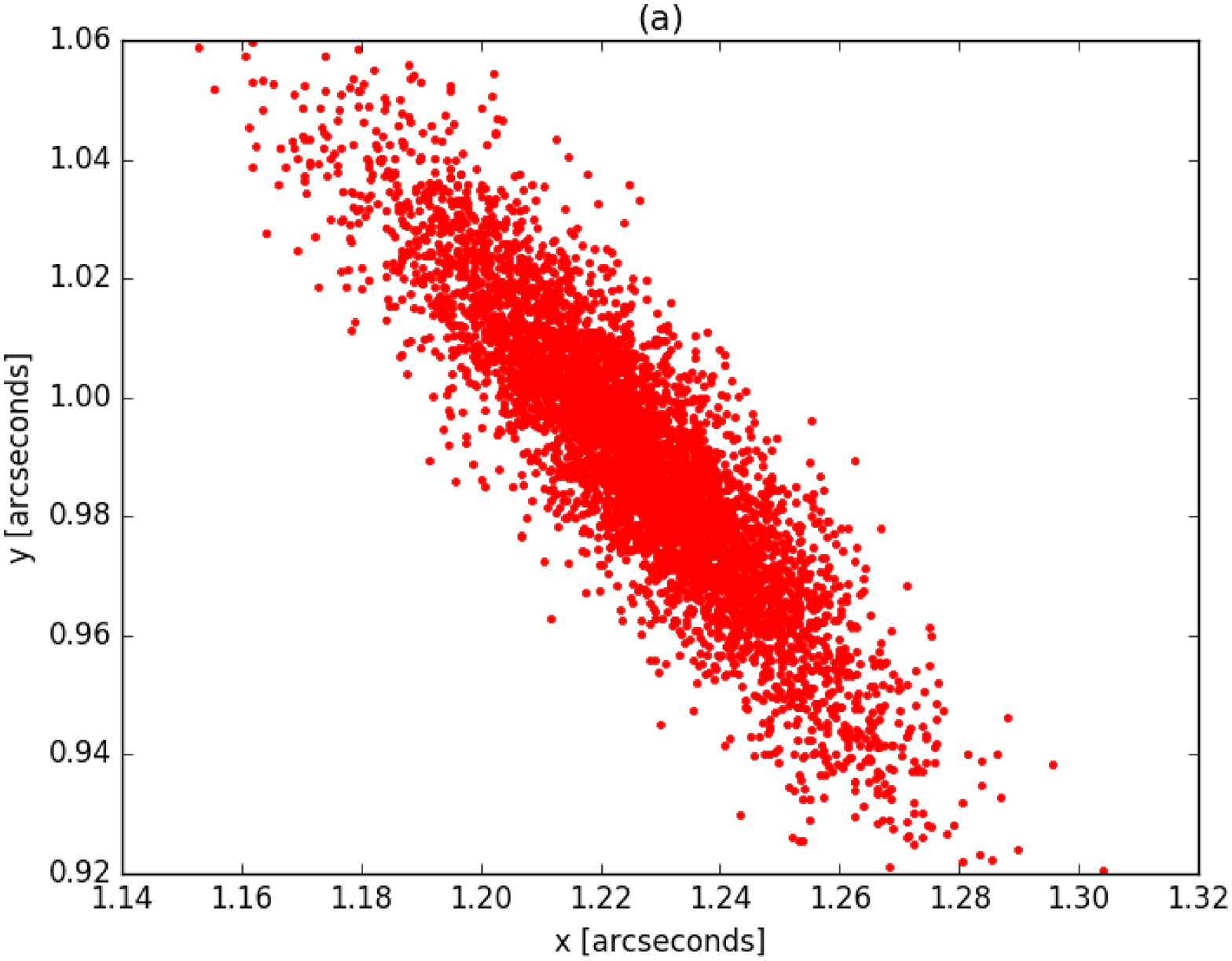}
 \includegraphics[width=8cm,angle=0]{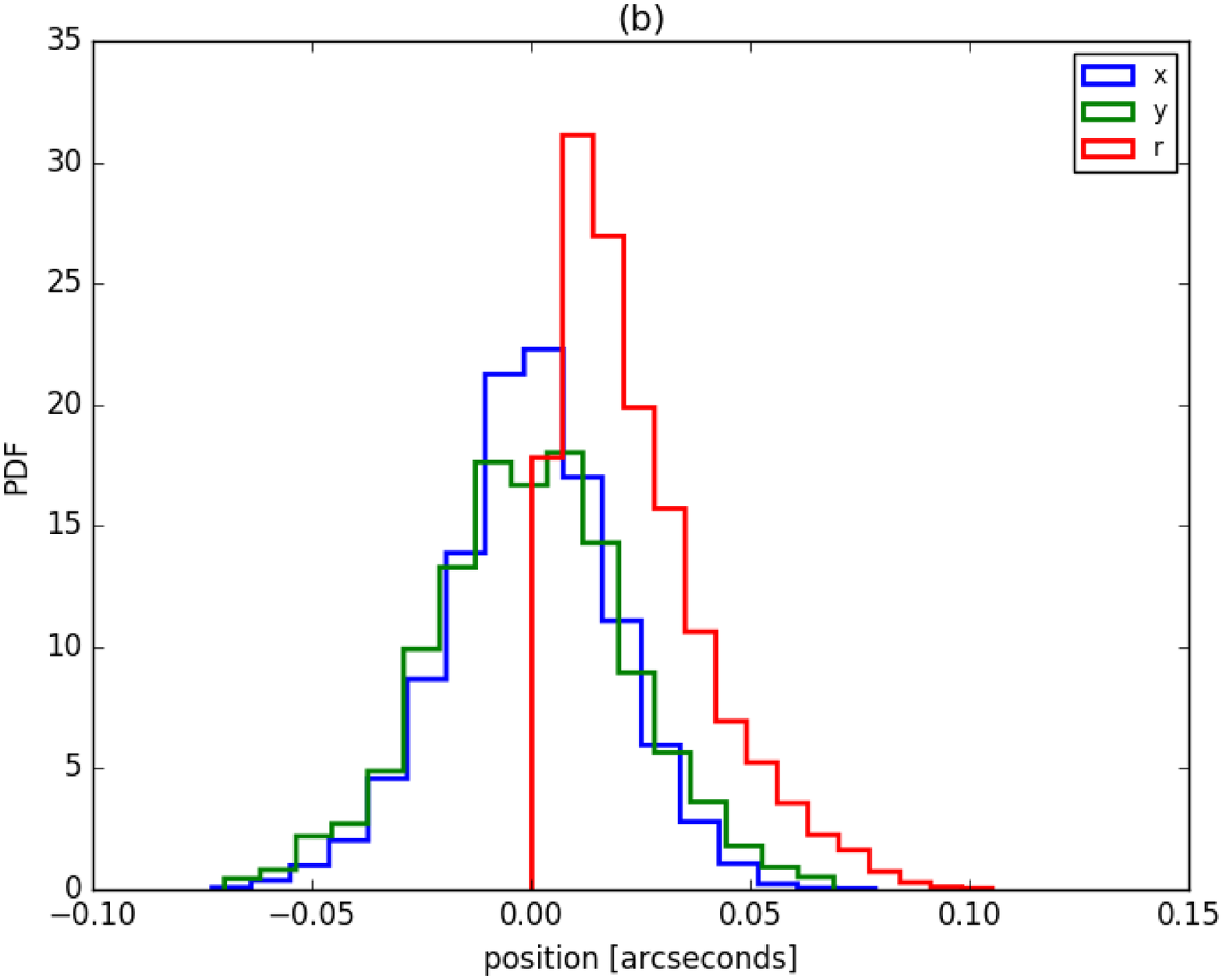}
  \caption{Position perturbation for image B in double-image system in case 1. (a) realizations from the same mass function,
  the orientation is the same as the local arc; (b) x, y and distance perturbations.
  }\label{positionfig}
\end{figure}

We simulated $10^4$ realizations of lensing sytems affected by dark matter subhalos by randomly choosing the subhalo positions.
For each realization, we calculated time delays between images. In rare cases, certain images might
split up. This phenomenon has been studied as an effect of small-scale structure. However, we focus on time delay anomalies and we ignored it.
We also noticed that for some cusp image systems, the order of arrival changed. Based on simulations we calculated the standard deviation of all perturbed time delays and treated it as $\sigma_{dm}$. Fig. \ref{results} (a) and (d) show the corresponding histograms.
Note that the image positions were also perturbed as shown in Fig. \ref{positionfig} and Tab. \ref{position}.

\subsection{Mock observations}
Simulation of images is based on the state-of-the-art H0LiCOW project standard~\citep{Ding2017}. We assumed that images are taken by the Hubble Space Telescope (HST) with the Wide Field Camera 3 (WFC3) IR channel in the F160W band. The corresponding PSF generated by {\tt Tinytim}\footnote{\url{http://www.stsci.edu/hst/observatory/focus/TinyTim}} is shown in Fig.~\ref{images} (f). Following common practice, we adopted eight dithered images and stacked them into a final image; the pixel size is $0.13''$ to $0.08''$, before and after drizzling. We added noise according to the realistic observation, including background, read noise and Poisson noise shown in Fig.~\ref{images} (d)  for double and (e) quad systems. Total exposure time was assumed as $1200s\times8 = 9600s$. For details of the simlulation, see \citep{Ding2017} and (arxiv/1801.01506).
Besides, we assume astrometric uncertainty as $0.005''$ and velocity dispersion uncertainty as $6.5\%$~\citep{Wong2017}.

\subsection{Best fits and $\sigma_{obs}$}
In order to fit the lens model to observations of arcs, velocity dispersion and image positions, we used a power law model with the radial profile $\rho(r)\propto r^{-\gamma}$ plus shear.
Note that in this process, we did not use the measurements of time delays from GWs. There were 17 free parameters consisting of source position (2), lens position (2), lens model parameters (6), host galaxy parameters (7). For different noise realizations, we fitted parameters of the lens and source positions to find the best smooth lens model and then we inferred corresponding time delays. To avoid local trapping of free parameters during optimalization process, we randomized the initial model parameter values, though it would take a longer time
to find the global minimum of the objective function.
The posterior PDF of model parameters $\boldsymbol{\xi}$ mentioned above can be expressed as:
\begin{equation}
P(\boldsymbol{\xi}|\mathbf{I},\sigma_v,\boldsymbol{\theta})\propto P(\mathbf{I},\sigma_v,\boldsymbol{\theta}|\boldsymbol{\xi})P(\boldsymbol{\xi}),
\end{equation}
where $\mathbf{I}$ stands for the pixel intensities of arcs, $\boldsymbol{\theta}$ stands for the point image positions.
The likelihood can be further written as a product of Gaussian distributions:
$P(\mathbf{I})P(\sigma_v)P(\boldsymbol{\theta})$, since these observations are independent.
For more details, we refer to the Bayesian analysis in~\citep{Suyu2013}.
We did not consider the lens light due to limitations of the glafic software~\citep{Oguri2010}. This would not change
the result much according to the H0LiCOW experience since the elliptical galaxy can be well modelled by a S$\acute{e}$rsic light model.

\begin{figure*}
 \centering
 \includegraphics[width=5cm,angle=0]{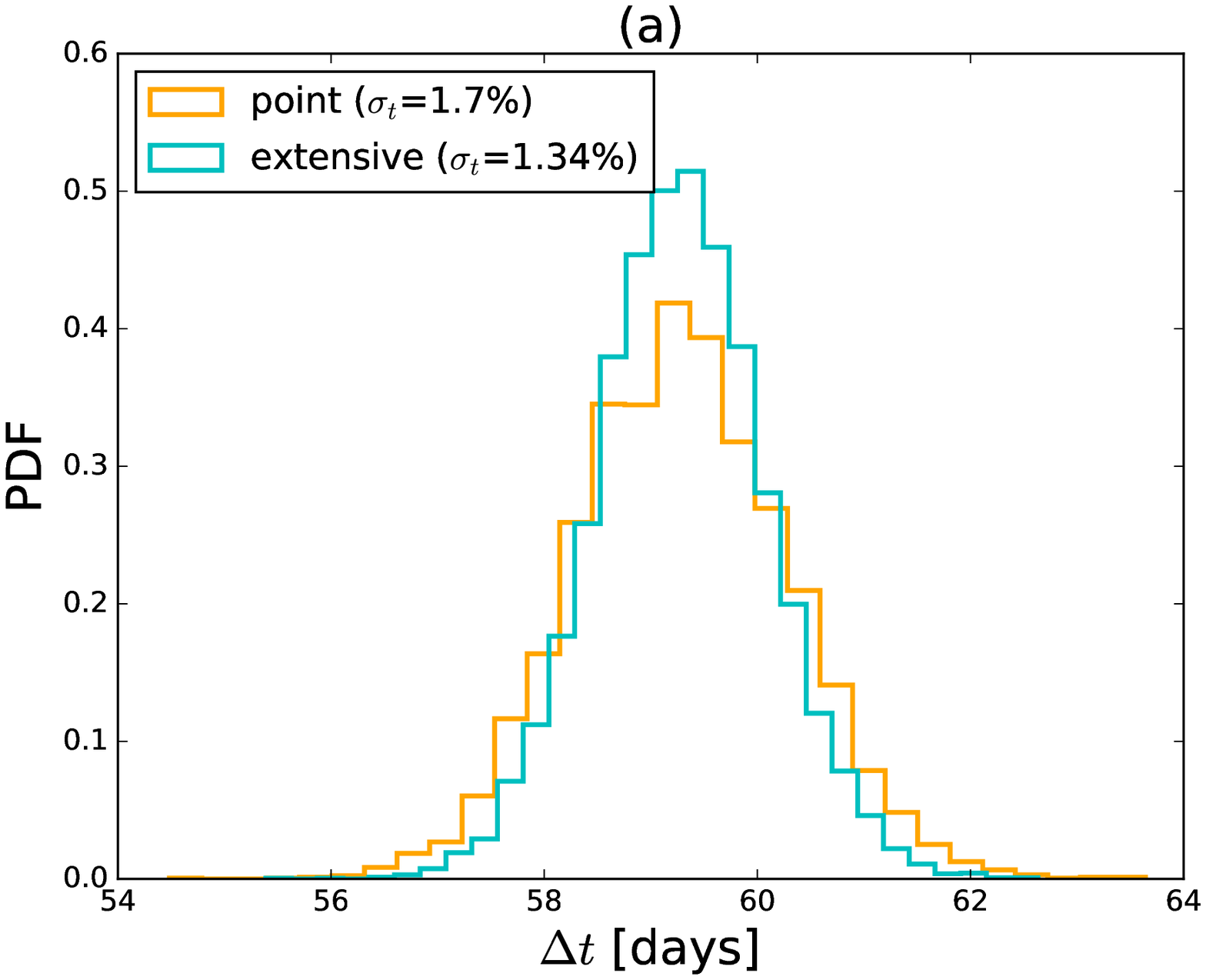}
 \includegraphics[width=5cm,angle=0]{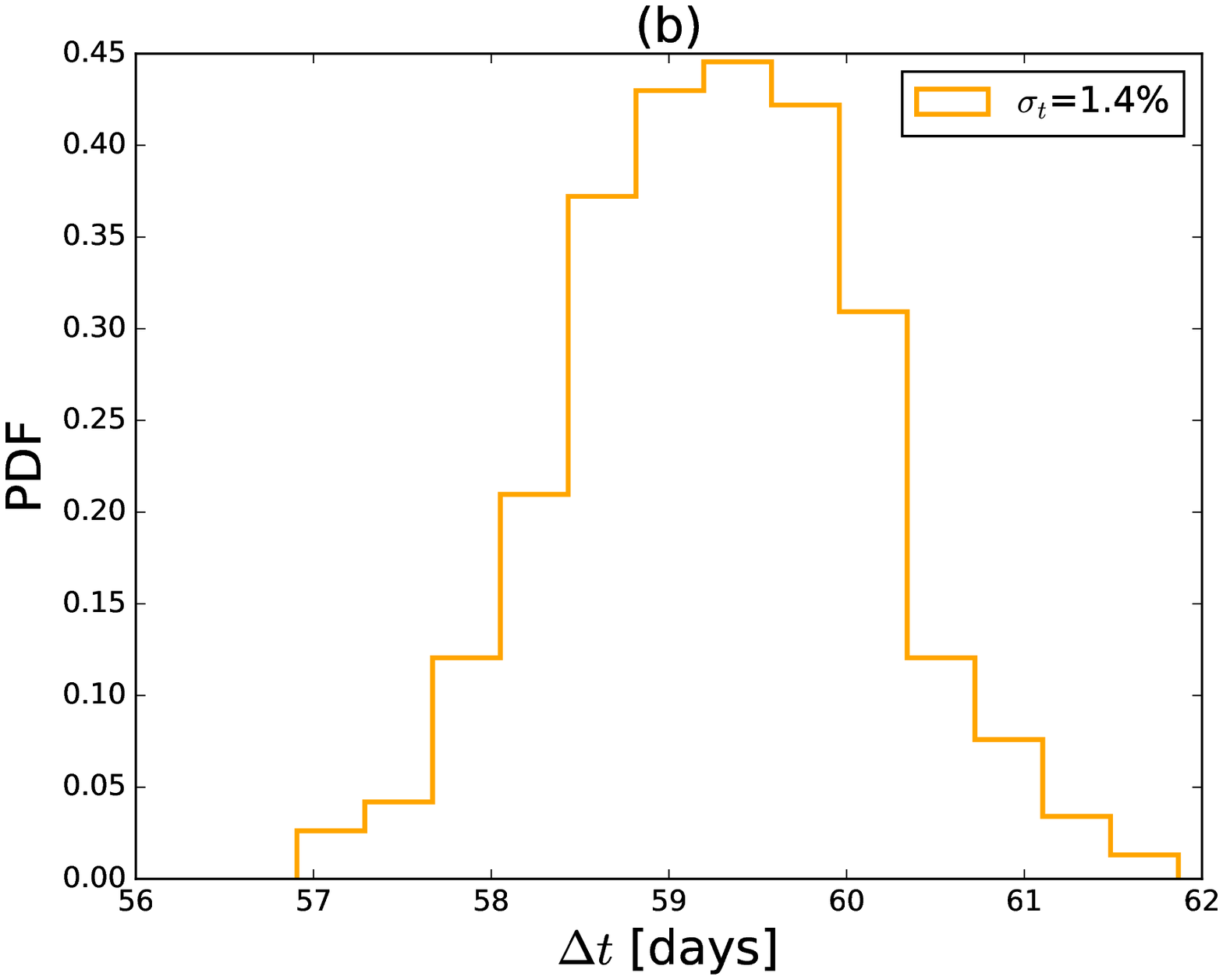}
 \includegraphics[width=5cm,angle=0]{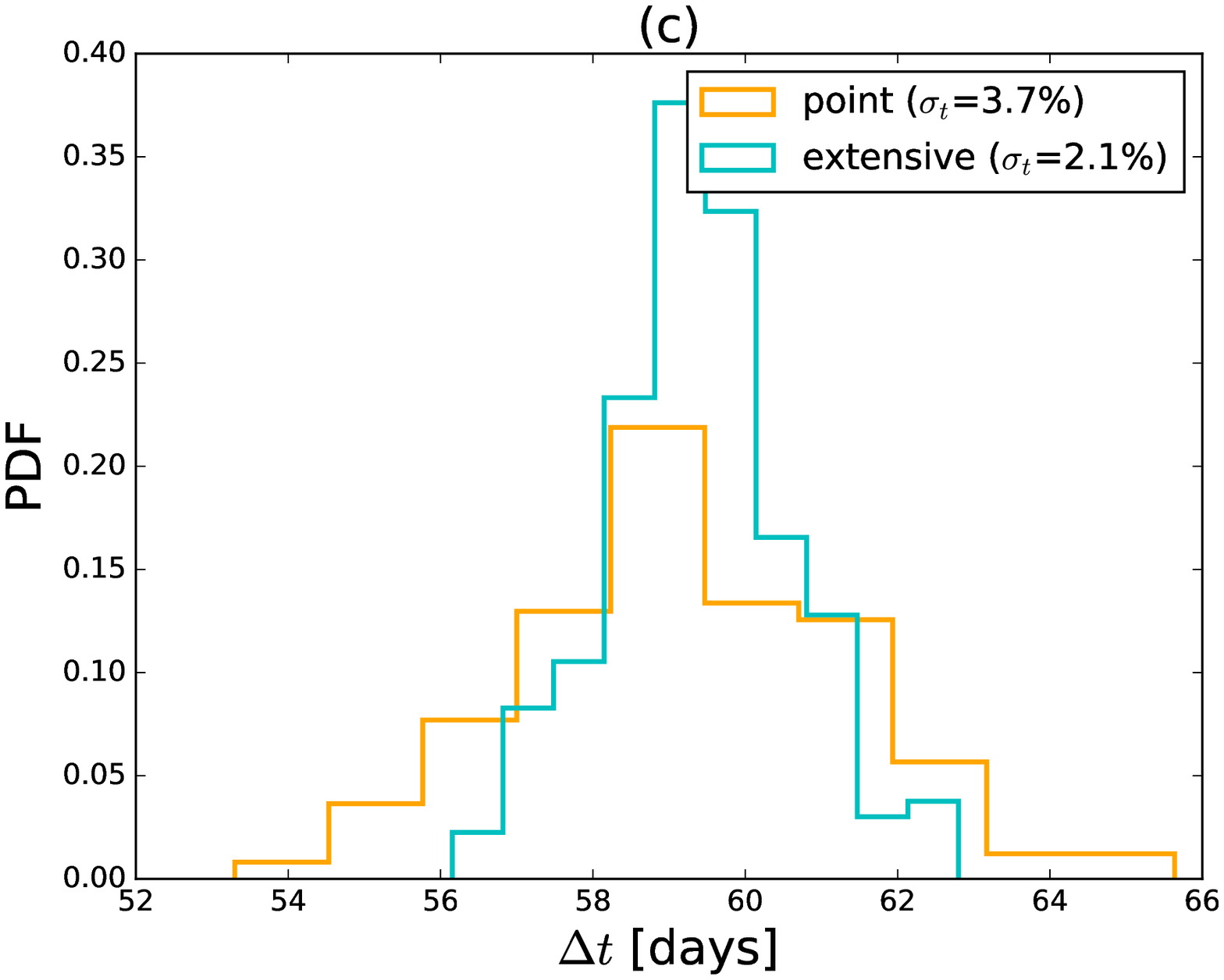}
 \includegraphics[width=5cm,angle=0]{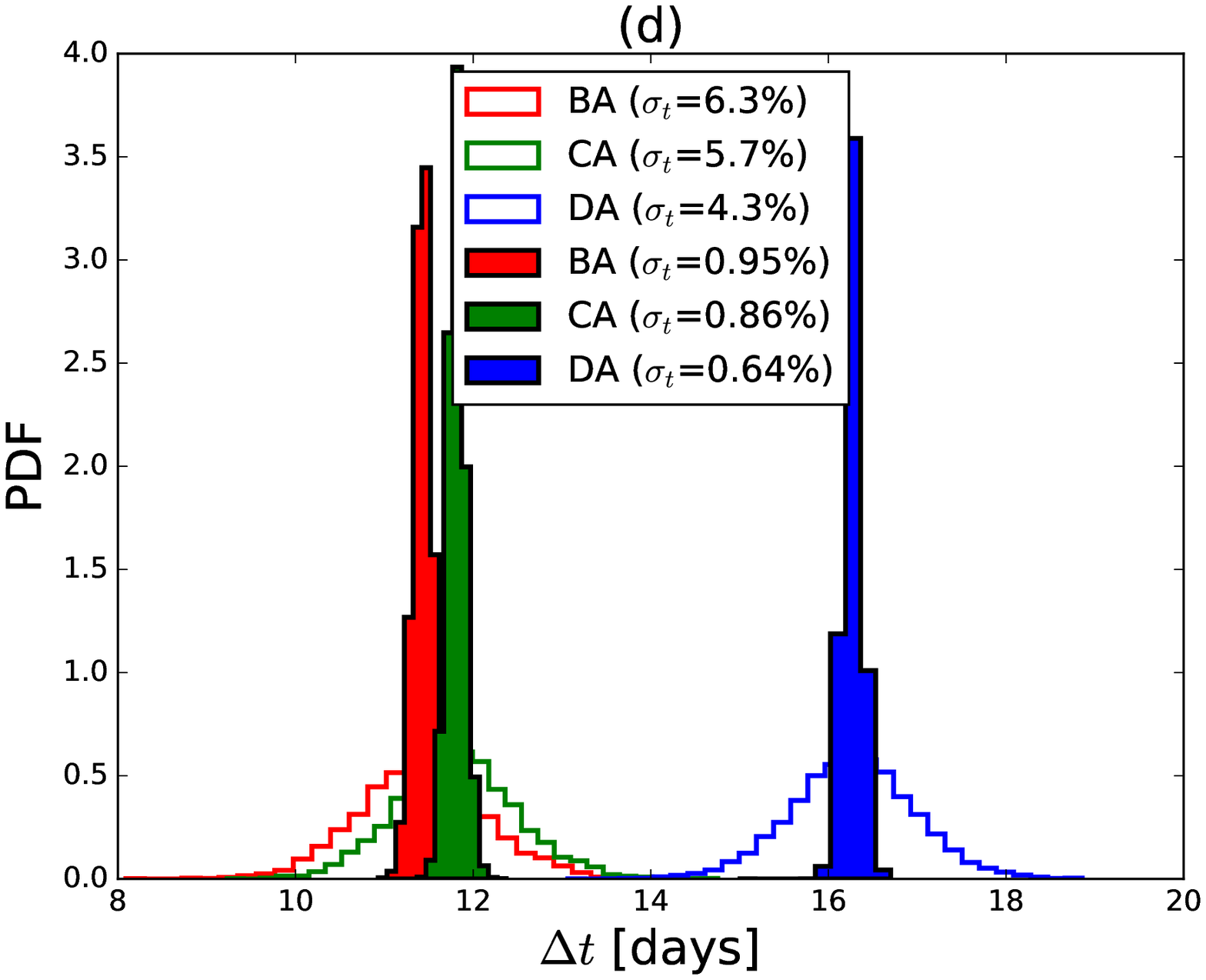}
 \includegraphics[width=5cm,angle=0]{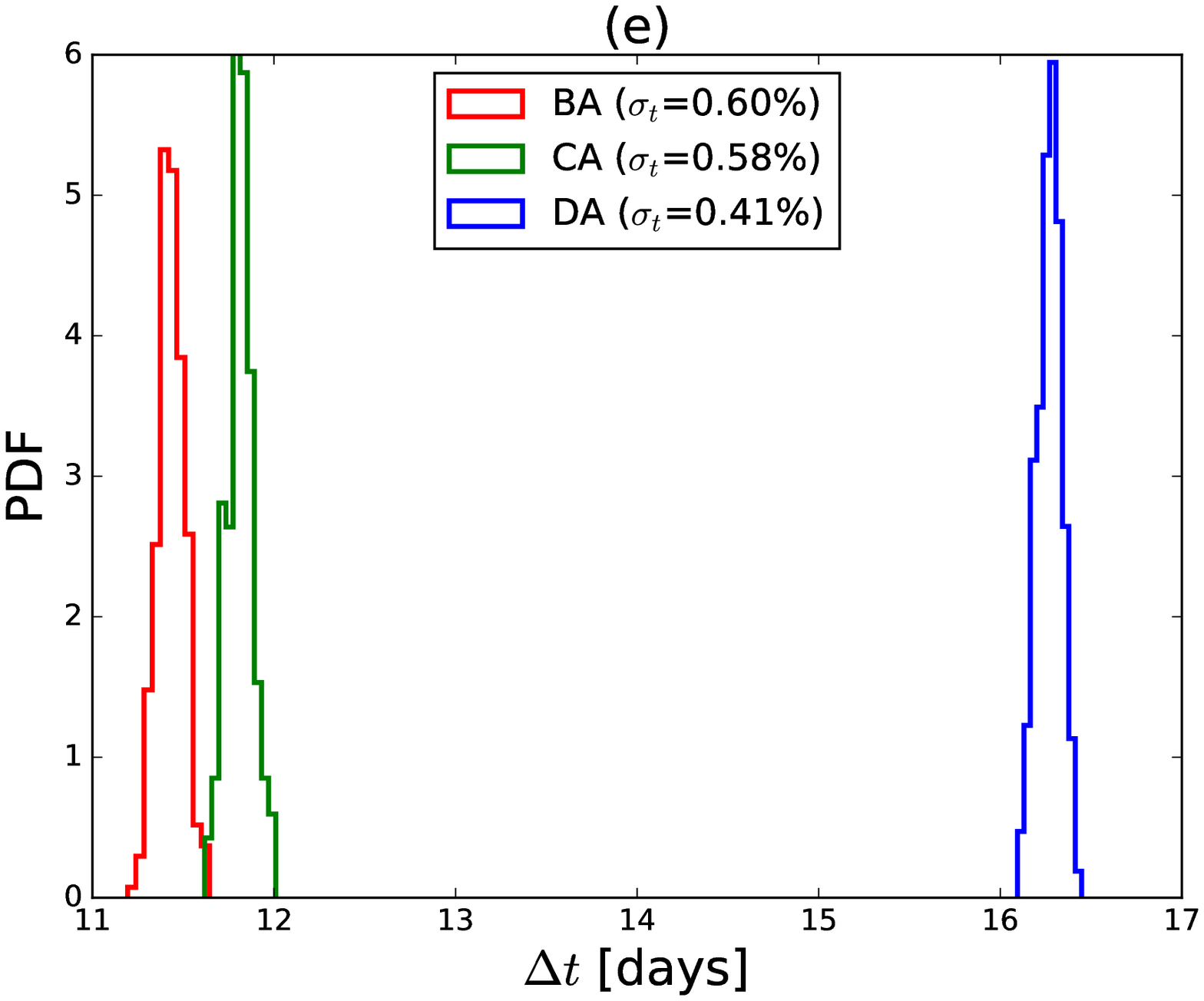}
 \includegraphics[width=5cm,angle=0]{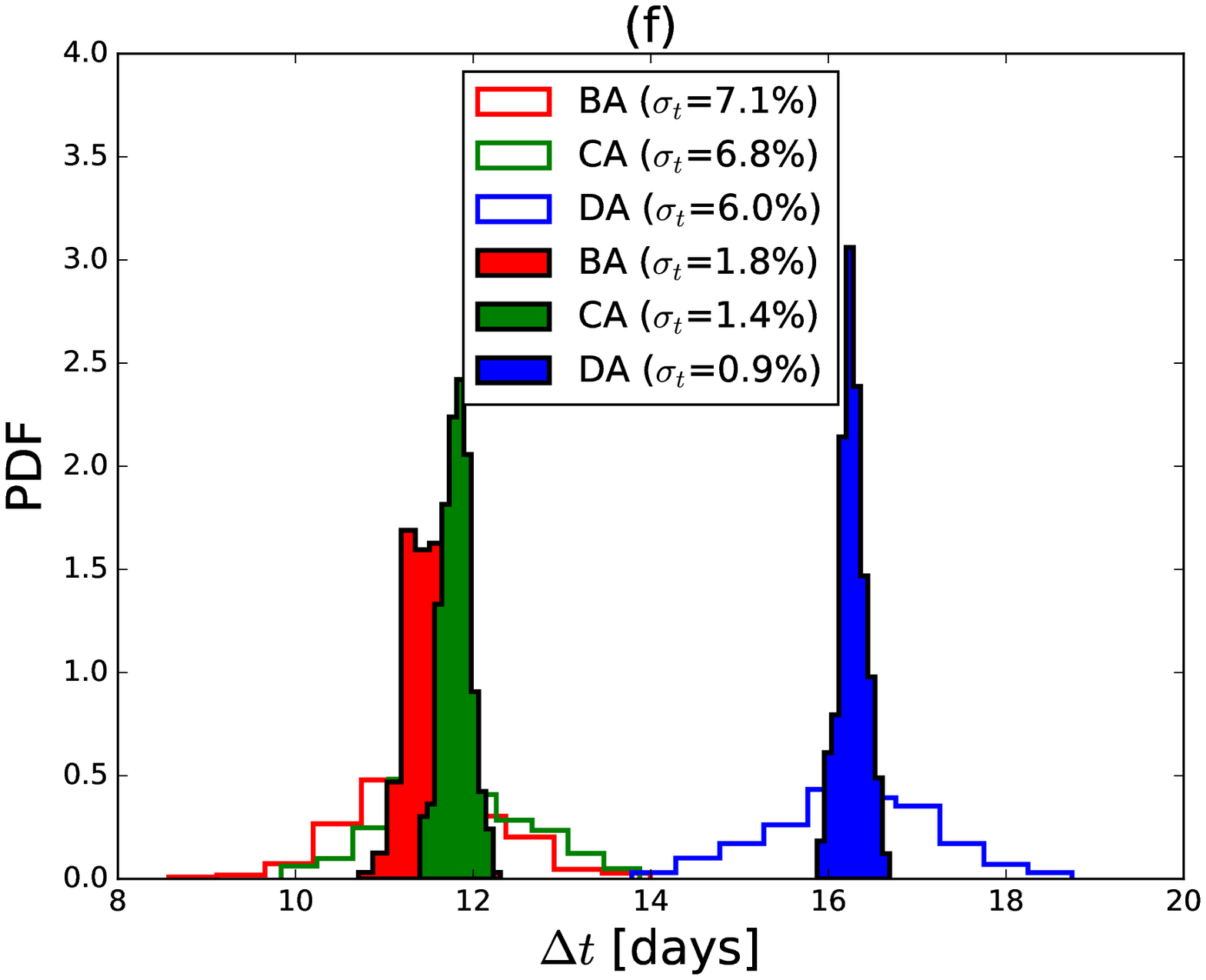}
  \caption{Uncertainty results for different mass functions and internal structures case for double and quad systems in case 1, 2, 3. (a) $\sigma_{dm}$ for double in case 1, 2;
  (b) $\sigma_{obs}$ for double; (c) $\sigma_{merge}$ for double in case 1, 2; (d) $\sigma_{dm}$ for quad in case 2, 3; (e) $\sigma_{obs}$ for quad; (f) $\sigma_{merge}$ for quad in case 2, 3;
  }\label{results}
\end{figure*}

We repeated this process 300 times by using different noises added to the observations, i.e., different realizations of the observations.
Finally, we calculated time delays based on the best fitted smooth lens model  and calculated the standard deviation of all time delays.
The histograms are shown in Fig. \ref{results} (b) and (e).

Note that in principle, to infer $\sigma_{obs}$, one should perform a MCMC simulation to get the
uncertainties for one noise realization based on perturbed arcs and image positions rather than a smooth
model. However, we just need to estimate the scale of $\sigma_{obs}$ rather than model a specific lensing system. Therefore
the average perturbation could be close to zero and the procedure we adopted is reasonable.

\begin{table*}\centering
\small
 \begin{tabular}{lccccccccccccc}
  \hline\hline
  Double &A(1)&A(2)&B(1) &B(2)& Quad &A(2)&A(3)& B(2)& B(3)& C(2)& C(3)& D(2)& D(3) \\
  \hline
 $r('')$ & 0.021& 0.018& 0.040& 0.037& &0.025& 0.004& 0.021& 0.004& 0.019& 0.003& 0.025& 0.005  \\
  \hline\hline
 \end{tabular}
 \caption{Position distance perturbations for all images in double and quad systems. x and y direction perturbations
 are Gaussian-like, and $r=\sqrt{x^2+y^2}$ is non-Gaussian, we show the values corresponding to the maximum probabilities.
}\label{position}
\end{table*}

\subsection{Host arc perturbation and $\sigma_{arc}$}
Dark matter subhalos change the lensing potential and perturb the observed host arcs. When we fit the arcs based on a certain smooth model, the corresponding systematic errors would occur.
To explore the scale of $\sigma_{arc}$, we used a similar approach like  in $\sigma_{obs}$. However, the noise map was added to the perturbed arcs, and then we fitted the smooth model. The resulted uncertainty $\sigma_{merge}$ should include both
$\sigma_{obs}$ and $\sigma_{arc}$ and the fitting $\chi^2$ should be larger than 1 due to unconsidered systematics. Actually, we notice that when studying cosmology, pixel uncertainties
were increased so that the $\chi^2=1$~\citep{Suyu2013} to avoid systematics.

Since $\sigma_{obs}$ and $\sigma_{arc}$
are merged together as $\sigma_{merge}$, while the former one is from observational noise, the latter one is from the mismatch between perturbed arcs and a smooth fitting model,
we must deduce $\sigma_{arc}$ from $\sigma_{obs}$ through $\sigma_{arc}^2$ = $\sigma_{merge}^2-\sigma_{obs}^2$.
We also investigated the covariance between $\sigma_{dm}$ and $\sigma_{arc}$ and found that it could be neglected.
Local subhalos near the point images primarily affect $\sigma_{dm}$, whereas $\sigma_{arc}$ is affected by all subhalos. They are approximately independent.

We summarize all relative uncertainties in Tab. \ref{doubledata} for the double image system and Tab. \ref{quaddata} for the quad image system.

\begin{table*}\centering
 \begin{tabular}{lcccccc}
  \hline\hline
  &$\sigma_{dm}$& $\sigma_{obs}$ & $\sigma_{merge}$ & $\sigma_{arc}$ & $\sigma_{stats}$ & $\sigma_{sys}$\\
  \hline
  case 1 & $1.7\%$ & $1.4\%$ & $3.7\%$ & $3.4\%$  & $2.64\%$ & $4.75\%$\\
  case 2 & $1.34\%$ & $1.4\%$ & $2.1\%$ & $1.57\%$ & $2.64\%$ & $2.30\%$ \\
  \hline\hline
 \end{tabular}
 \caption{Uncertainties for double image system for case 1 and 2.
}\label{doubledata}
\end{table*}

\begin{table*}\centering
 \begin{tabular}{lccccccc}
  \hline\hline
  & &$\sigma_{dm}$& $\sigma_{obs}$ & $\sigma_{merge}$ & $\sigma_{arc}$ & $\sigma_{stats}$ & $\sigma_{sys}$\\
  \hline
  case 2 & BA & $6.3\%$ & $0.60\%$ & $7.1\%$ & $7.07\%$  & $2.3\%$ & $9.47\%$\\
    & CA  & $5.7\%$ & $0.58\%$ & $6.8\%$ & $6.78\%$  & $2.3\%$ & $8.85\%$\\
     &DA  & $4.3\%$ & $0.41\%$ & $6.0\%$ & $5.99\%$  & $2.27\%$ & $7.37\%$\\
  case 3 & BA & $0.95\%$ & $0.60\%$ & $1.8\%$ & $1.7\%$  & $2.3\%$ & $1.94\%$\\
    & CA  & $0.86\%$ & $0.58\%$ & $1.4\%$ & $1.27\%$  & $2.3\%$ & $1.54\%$\\
     &DA  & $0.64\%$ & $0.41\%$ & $0.9\%$ & $0.8\%$  & $2.27\%$ & $1.03\%$\\
  \hline\hline
 \end{tabular}
 \caption{Uncertainties for quad image system for case 2 and 3.
}\label{quaddata}
\end{table*}

\section{Discussion}

Study of dark matter substructure using time delay anomaly method is unaffected by dust extinction and stellar microlensing.
We proposed to use lensed GW signals accompanied by electromagnetic counterparts to enhance the performance of this
approach, which will make it a promising and robust probe.

The very term "time delay anomaly" tacitly suggests the following procedure: fit part of the observational data with a smooth lens model and then compare the inferred time delay with the measured one. If the difference between the two is larger than acceptable (conservative)
statistical uncertainty -- anomalies occur.

Previous works based on lensed quasars used the information encoded in lensed images assuming a point source to infer the smooth model via
Monte Carlo simulation based on certain galaxy catalogs. Moreover, time delay measurements from the light curves had considerable uncertainties. Consequently, the statistical uncertainties were
too large comparing with the systematical uncertainty from dark matter. Therefore it was hard to identify the effects of dark matter halos
in a reliable manner.
As for the anomalies found in RX J1131-1231 and B1422+231~\citep{Congdon2010}, one could expect that they might be caused by other systematics rather than the dark matter subhalos described in ref.~\citep{Keeton2009}.
Besides, in previous methods, only quad systems with time delay ratio measurements could be used robustly, due to the radial mass profile degeneracy. The power law
slope index could affect time delay between two images, but the time delay ratio should be immune to it.

The advantage of the method we propose is that while the lensed quasar could only measure time delay at the percent level through sampled light curves, the lensed GW signal provides a very accurate measurement due to its transient nature. With the improvement of the quality of
optical images, we propose to directly extract information about the smooth model from the arcs.
Without contamination of the host galaxy image by bright AGN, the lensed GW system identified in the optical, could provide complete host arcs.
This would contribute to lens modelling, thus decreasing considerably  the value of $\sigma_{obs}$ as part of statistical uncertainties. Then one might be able to uncover systematical uncertainties more easily.
Our method could test both double and quad systems since it directly fits the lens parameters like the slope parameter to the images and contains no slope-time delay degeneracy.
For cusp images, the image order method should be more robust.
We also considered the statistical uncertainty brought by cosmology and line of sight fluctuations.

As one can see from Tab. \ref{doubledata} and Tab. \ref{quaddata}, under the assumptions we made,
statistical uncertainties are
$\sim 2.5\%$ and systematical uncertainties range from $1\%$ to $10\%$, depending on the dark matter
subhalo mass functions, internal structure and lensing configurations.
We conclude that systematical uncertainties are comparable to
statistical ones, which is very promising.
Concerning lensed quasars, $\sigma_{obs}$ could be several times larger~\citep{Liao2017} due to bright AGNs,,
and one needs also to consider statistical uncertainties from time delay measurements via sampled light curves. These may lead to very large statistical uncertainties compared with
systematical ones and in most cases would make dark matter subhalo effects hard to probe.

We emphasize that in this work we only discussed the statistical relationship between mass function and perturbation uncertainties. Actually, one can not get a quantitative result from the measurement of a single system. However, each measurement could provide
the lower limit of the substructure according to measured time delay anomalies. With more available systems, dark matter
substructure will be assessed more accurately.

On the other hand, we also notice that in many cases, just one or a few subhalos could explain the observed anomalies. For example,
the satellite in lens RXJ1131-1231~\citep{Suyu2013} and the subhalo in lens B1422+231~\citep{Nierenberg2014} could fit the observed flux
ratios well. For such cases, lensed GW systems would be more powerful and probe these substructures more accurate.

\textbf{Throughout this work, we assumed that the dark matter substructure is solely responsible for time delay anomalies.
However, there are known cases where the large-scale substructure is in the form of disks~\citep{Hsueh2016,Hsueh2017,Hsueh2018}.
When the disks are edge-on oriented, they can also generate flux-ratio anomalies and time delay anomalies as well.
Therefore, one should be very careful about the complexity of baryonic structure. It is critical to directly
detect the edge-on disks or massive luminous satellites through high-resolution imaging. For example, the Keck adaptive optics or HST imaging may reveal these disks. One may also try to constrain the mass of the disk independently, through kinematic measurements. On the other hand, we need further simulations and emulations to see whether these baryonic structures could be distinguished.
Certainly, more complex mass models will have to be considered for a robust quantification of dark matter substructure.
In this context, we also refer to the ongoing TDLMC program whose goal is to assess the present capabilities of lens modeling codes and assumptions and test the level of accuracy of cosmological inference. In this program, the team generating mock data will add some systematics to see whether the community could recover them with state-of-the art modeling techniques.
}

\section{Perspectives}
We have demonstrated that the lensed GW signal accompanied by the electromagnetic counterpart could be an excellent tool to study
dark matter substructure in galaxies by its accurate time delay measurements.
When combined with flux ratio, astrometric and small-scale structure measurements, the dark matter subhalo mass function could be tested.
In particular, flux ratios should not be confounded by microlensing due to the long wavelengths of GWs.
It is different from the traditional radio loud quasars whose source sizes are extended enough for convolution of the magnification map.
The limitation of this method lies in relatively small number of such systems supposed to be detected by
third-generation GW detectors, i.e. 2-10 per year for binary neutron stars or neutron-black hole systems with electromagnetic counterparts.
The methodology we presented can be also applied in lensed  quasar systems, with the caveats discussed above.
As for the lensed quasars, the upcoming LSST will bring us $\sim 400$ systems with well-measured time delays in 10 years~\citep{Liao2015}.
One may combine many such lensed quasar systems to improve the measurement precision. However, in such case one should make some statistical assumptions like that the dark matter substructure in all  lensed systems is similar. On the other hand,
with lensed GW signals individual lenses could be probed better. We look forward to seeing these systems detected and our method applied in studying dark matter substructure.

\section*{Acknowledgments}
K. Liao was supported by the National Natural Science Foundation of China (NSFC) No. 11603015
and the Fundamental Research Funds for the Central Universities (WUT:2018IB012).
X. Ding acknowledges support by China Postdoctoral Science Foundation Funded Project (No. 2017M622501).
M. Biesiada was supported by Foreign Talent Introducing
Project and Special Fund Support of Foreign Knowledge
Introducing Project in China. He also acknowledges hospitality of the Wuhan University.
This research was also partly supported by the Poland- China Scientific \&
Technological Cooperation Committee Project No. 35-4.
Z.-H. Zhu was supported by the National Basic Science Program (Project 973) of China under (Grant No. 2014CB845800), the National Natural Science Foundation of China under Grants Nos. 11633001 and 11373014, the Strategic Priority Research Program of the Chinese Academy of Sciences, Grant No. XDB23000000 and the Interdiscipline Research Funds of Beijing Normal University. X.-L. Fan was supported by NSFC No. 11633001, 11673008 and Newton International Fellowship Alumni Follow on Funding.

\clearpage

\end{document}